\documentclass{ptapap}

\author{Radosław Smolec}[smolec@camk.edu.pl,CAMK]
\author{Paweł Moskalik}[CAMK]
\author{Nancy R. Evans}[SAO]
\author{Anthony F. J. Moffat}[AFM]
\author{Gregg A. Wade}[GW]
\team{the BEST}[BEST]
\affil[CAMK]{Nicolaus Copernicus Astronomical Center, Bartycka 18, 00--716 Warszawa, Poland}
\affil[SAO]{Smithsonian Astrophysical Observatory, MS 4, 60 Garden Street, Cambridge, MA 02138, USA}
\affil[AFM]{D\'epartement de physique, Universit\'e de Montr\'eal, C.P. 6128, Succursale Centre-Ville, Montr\'eal, Qu\'ebec, H3C 3J7, Canada}
\affil[GW]{Dept. of Physics, Royal Military College of Canada, PO Box 17000, station Forces, Kingston, Ontario, Canada K7K 4B4}
\affil[BEST]{BRITE Executive Science Team}

\title{BRITE Observations of Classical Cepheids -- an Update}

\begin{document}

\maketitle

\begin{abstract}
We briefly summarize the BRITE observations of classical Cepheids so far.
\end{abstract}

\section{Introduction}

BRITE is a constellation of five satellites, each equipped with a 3-cm telescope, observing the brightest stars in the sky \citep[see e.g.,][]{weiss,pablo}. Its primary targets are stars brighter than $4$\thinspace mag in $V$-band, however fainter stars, with slow variability, such as classical Cepheids, can be observed at high enough precision for their normally large amplitudes. So far BRITE has observed eleven Cepheids. Initial results were reported in \cite{brite2} and here we provide a short update.

Table~\ref{tab} provides basic data on the observed Cepheids, of which six pulsate in the fundamental mode (top section of the Table) and five pulsate in the first overtone (bottom section). The consecutive columns provide the star's name and HD number, period as determined from the analysis of BRITE data, mean $V$-band brightness and a summary of observational data, i.e. satellite ID (UBr -- UniBRITE, BTr -- BRITE Toronto, BHr -- BRITE Heweliusz) and field ID. A short series of test observations in the blue filter are not included.

\begin{table}
\begin{tabular}{lrrrl}
star & HD & $P$\thinspace (d) & $\langle V\rangle$& summary of observational data\\
\hline
\multicolumn{5}{l}{\it fundamental mode Cepheids:}\\
T~Vul        & 198726 &  4.4355 & 5.75 & UBr, BTr (Cyg-I), BTr (Cyg-II)\\
$\delta$~Cep & 213306 &  5.3663 & 3.95 & BHr, BTr (CasCep-I)\\
X~Sgr        & 161592 &  7.0128 & 4.55 & UBr (Sgr-I), BHr (Sgr-II)\\ 
W~Sgr        & 164975 &  7.5949 & 4.67 & UBr (Sgr-I), BHr (Sgr-II)\\ 
l Car        &  84810 & 35.601~ & 3.72 & UBr (Car-I)        \\     
U Car        &  95109 & 38.868~ & 6.29 & BHr (Car-I)        \\
\multicolumn{5}{l}{\it first overtone Cepheids:}\\
DT~Cyg       & 201078 &  2.4991 & 5.77 & UBr, BTr (Cyg-I), BTr (Cyg-II)\\
V1334~Cyg    & 203156 &  3.3328 & 5.87 & UBr, BTr (Cyg-I), BTr (Cyg-II)\\
BG Cru       & 108968 &  3.3427 & 5.49 & BTr (CruCar-I)        \\
AH Vel       &  68808 &  4.2272 & 5.70 & BHr, BTr (VelPic-II)  \\
MY~Pup       &  61715 &  5.6953 & 5.68 & BHr (VelPic-I) BHr, BTr (VelPic-II)\\
\hline
\end{tabular}
\caption{Basic data about Cepheids observed with BRITE (pulsation period, mean $V$-band brightness) and indication of satellites and observing campaigns in which data were gathered (red filter only; see \textsf{http://brite.craq-astro.ca/} for more details). Stars are sorted by pulsation mode (fundamental mode stars in the top section of the table) and by increasing pulsation period.}
\label{tab}
\end{table}

The data reduction procedure was briefly described in \cite{brite2}. In a nutshell, the variability was first modelled with a Fourier series. The residuals were used to decorrelate the data with parameters like CCD temperature, star's position on the CCD or the satellite's orbital phase. The decorrelated data were then orbit-averaged. The remaining slow residual trends were modelled with polynomials. Outliers were removed during the procedure. Further details on BRITE photometry and its reduction can be found in \cite{popowicz}.

\section{Results}

\begin{figure}[t]
\includegraphics[width=\textwidth]{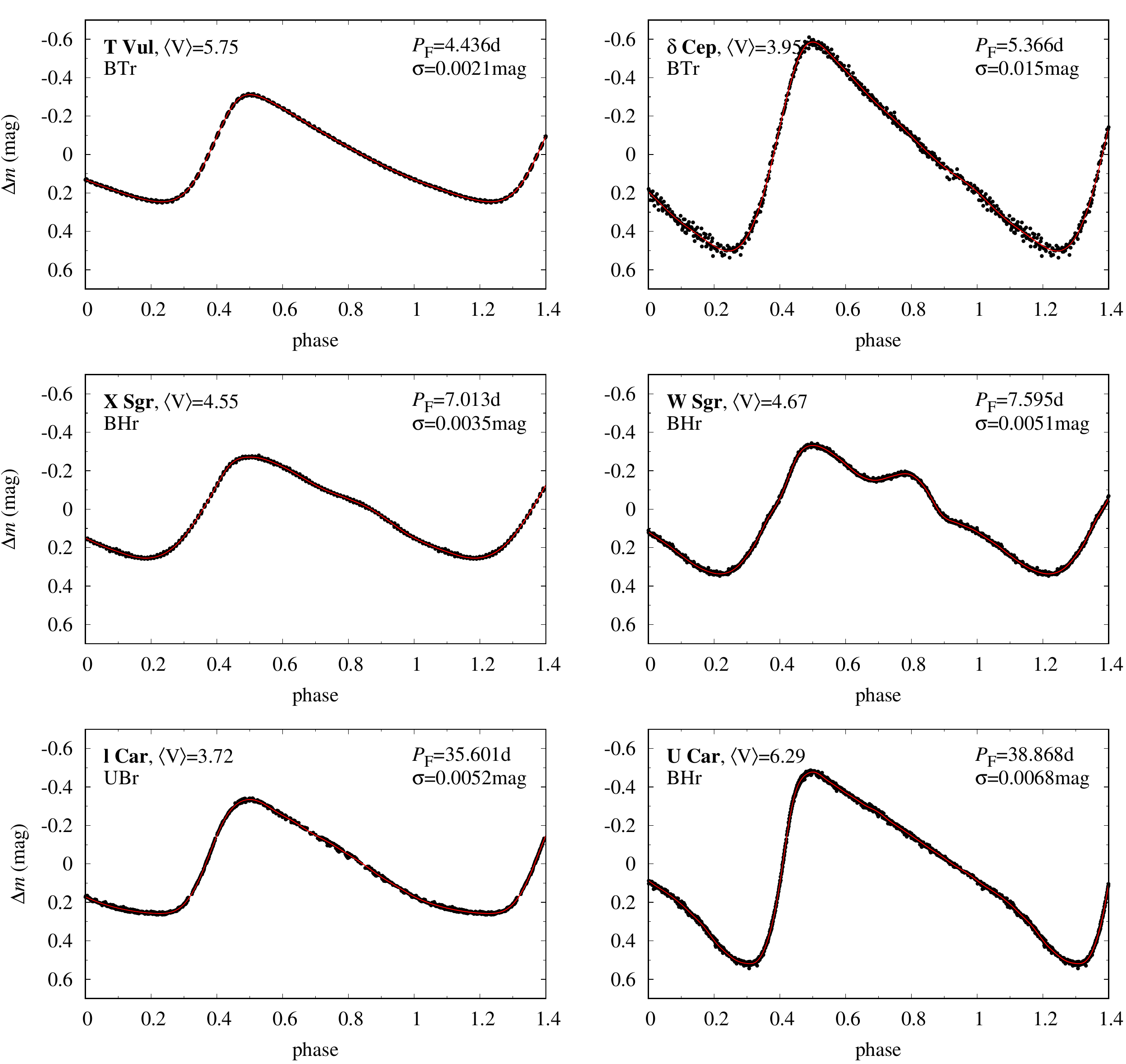}
\caption{Phased light curves with Fourier fits for the fundamental mode Cepheids observed with BRITE-Constellation. Basic data about the Cepheids is given in each panel.}
\label{fig:lcF}
\end{figure}

\begin{figure}[t]
\includegraphics[width=\textwidth]{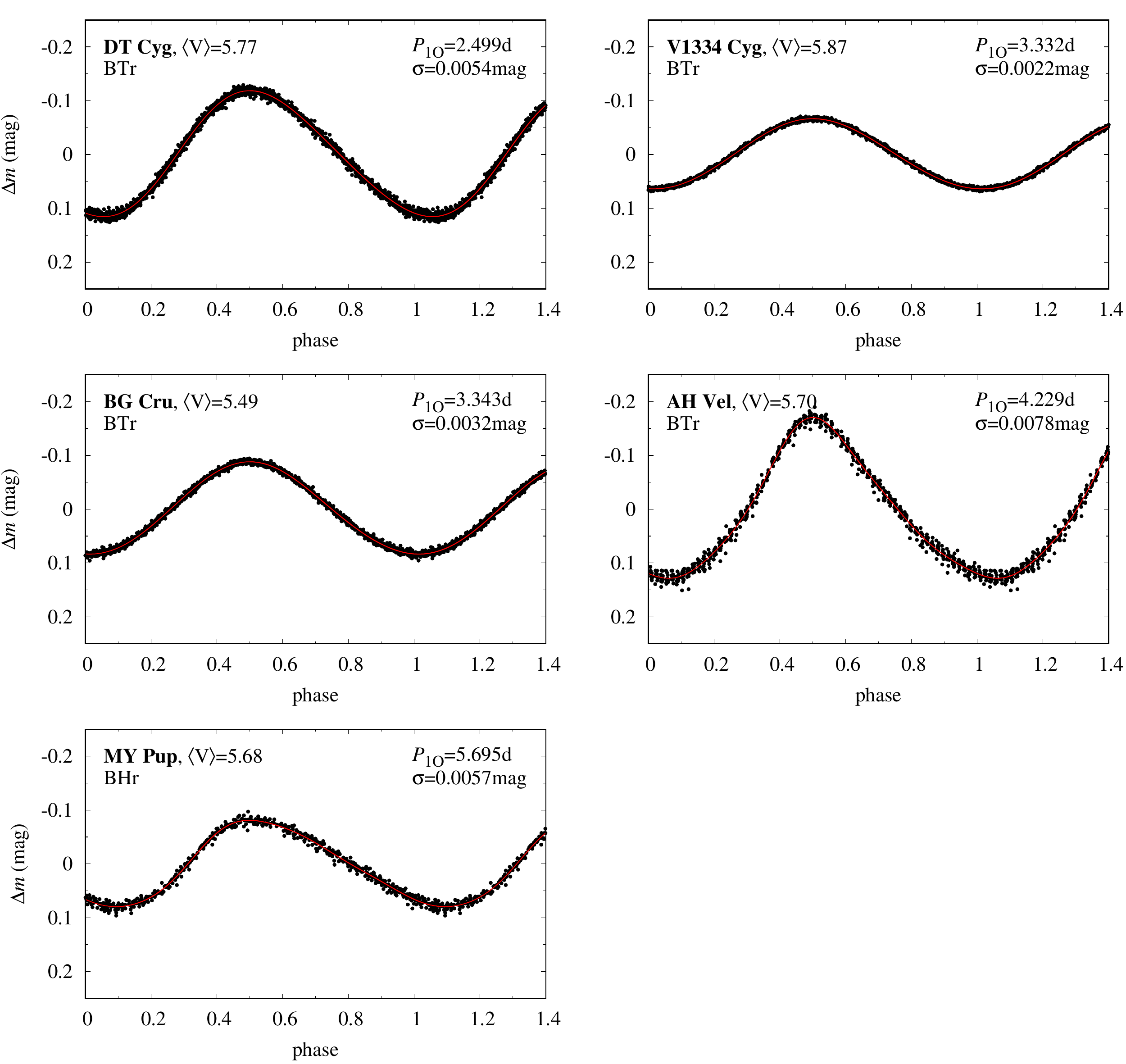}
\caption{Phased light curves with Fourier fits for the first overtone Cepheids observed with BRITE-Constellation. Basic data about the Cepheids is given in each panel.}
\label{fig:lcO}
\end{figure}

Figs.~\ref{fig:lcF} and \ref{fig:lcO} present a gallery of phased light curves for the fundamental mode and first overtone Cepheids, respectively. In each figure, stars are ordered with the increasing pulsation period. Data are plotted along with the Fourier fits; the data dispersion is given in each panel. Below we briefly summarize the most interesting results for each of the observed Cepheids, starting with the fundamental mode pulsators.

{\bf T~Vul}. No new data were gathered since our previous report. In the frequency spectrum, the radial fundamental mode is non-stationary. Time-dependent analysis indicates that a long-period modulation might be present in this star. Details in \cite{brite2}.

{\bf $\boldmath{\delta}$~Cep}. No new data were gathered since our previous report. Despite being one of the brightest Cepheids in our sample, the data are of the lowest quality and do not allow any detailed analysis. See the discussion in \cite{brite2}.

\begin{figure}[t]
  \centering
\includegraphics[width=.8\textwidth]{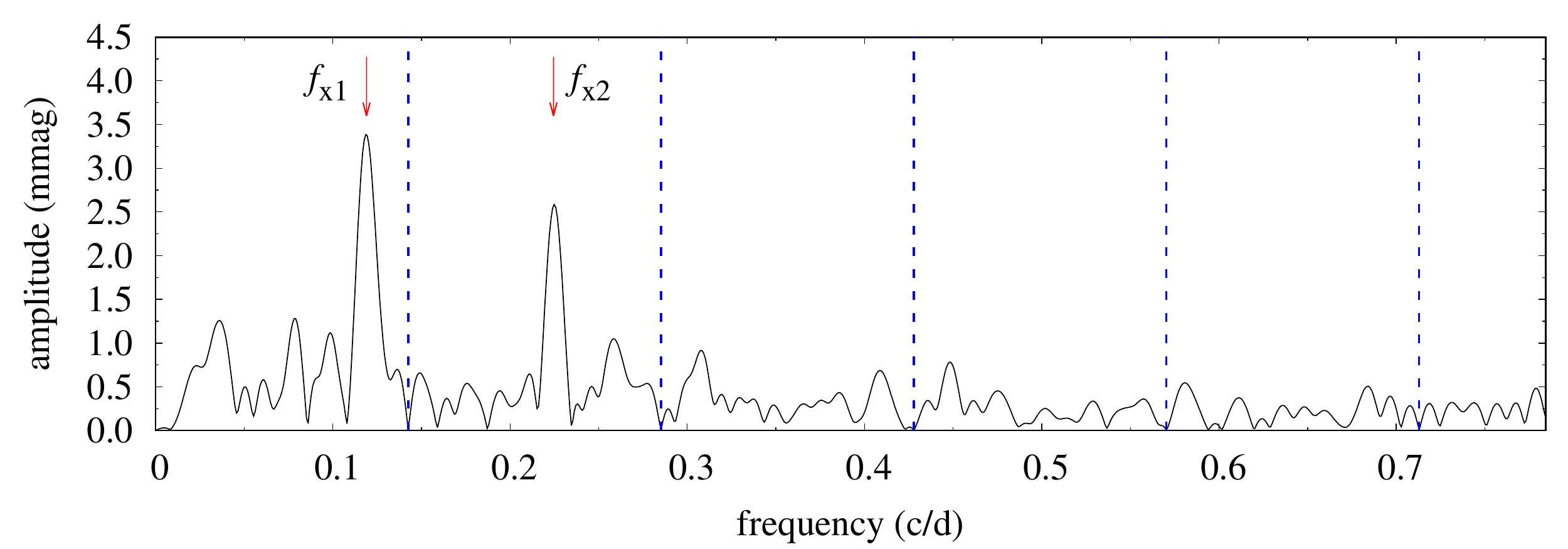}
\caption{Frequency spectrum for X~Sgr after prewhitening with the fundamental mode and its harmonics (dashed lines). Two additional signals, marked with arrows, are clearly detected.}
\label{fig:xsgr}
\end{figure}

{\bf X~Sgr} and {\bf W~Sgr}. The initial data for these stars covered only four pulsation cycles and were gathered by UBr. Its performance for faint stars is much worse than that of BTr and BHr. The new data were gathered by BHr and extend for more than 16 pulsation cycles, which allows a proper analysis of the frequency content. In the frequency spectrum of W~Sgr no additional signals are detected except the fundamental mode and its harmonics. X~Sgr appears much more interesting. In addition to radial fundamental mode ($f_0=0.142593$\thinspace c/d) two other periodicities were detected (see Fig.~\ref{fig:xsgr}): $f_{\rm x1}=0.1190$\thinspace c/d ($A_{\rm x1}=3.6$\thinspace mmag) and  $f_{\rm x2}=0.2246$\thinspace c/d ($A_{\rm x2}=2.8$\thinspace mmag). The corresponding period ratios are $P_{\rm x1}/P_0=1.20$ and $P_{\rm x2}/P_0=0.635$. These additional periods cannot correspond to radial overtones; their origin remains unknown. Possible hypotheses include: non-radial pulsation, low-amplitude periodic modulation, which was recently reported in several tens of fundamental mode Cepheids by \cite{cepmod} (interpretation for $f_{\rm x1}$), or contamination. Interestingly, X~Sgr and three first overtone Cepheids, V1334~Cyg, BG~Cru and EV~Sct (of which the first two were also observed by BRITE, see below) show unusual line profile structures, as investigated by \cite{kovtyukh}. They suggested that these features can be caused by the non-radial pulsations. The other interpretation are multiple shock waves propagating in the atmosphere \citep{mathias}.

{\bf l~Car} and {\bf U~Car}. These are the longest-period Cepheids in our sample. Only three pulsation cycles were observed for each of them. Despite U~Car being significantly fainter, data quality is similar to that for l~Car. This is because U~Car was observed by BHr, while l~Car by UBr, performance of which is much worse than BHr. For these two stars pulsation periods were determined directly from the data, by minimizing the dispersion of the Fourier fit. Small cycle-to-cycle differences are apparent in the light curve of l~Car. Interestingly, \cite{anderson} reported a cycle-to-cycle variability of spectral lines and atmospheric velocity gradients in this star. Without more observation and more detailed analysis however, we cannot rule out the possibility that the changes we observe in the BRITE data are of instrumental origin. The light curve of U~Car appears more stable.

{\bf DT~Cyg} and {\bf V1334~Cyg}. No new data were gathered for these first overtone pulsators since our previous report. Additional periodicities, that might be due to modulation or due to non-radial modes, were detected in both stars. The reader is referred to \cite{brite2} for more details.

{\bf BG~Cru}. Good quality data were gathered for BG~Cru by BTr. Its frequency spectrum contains no significant periodicities in addition to radial first overtone and its harmonic.

\begin{figure}[t]
  \centering
\includegraphics[width=.8\textwidth]{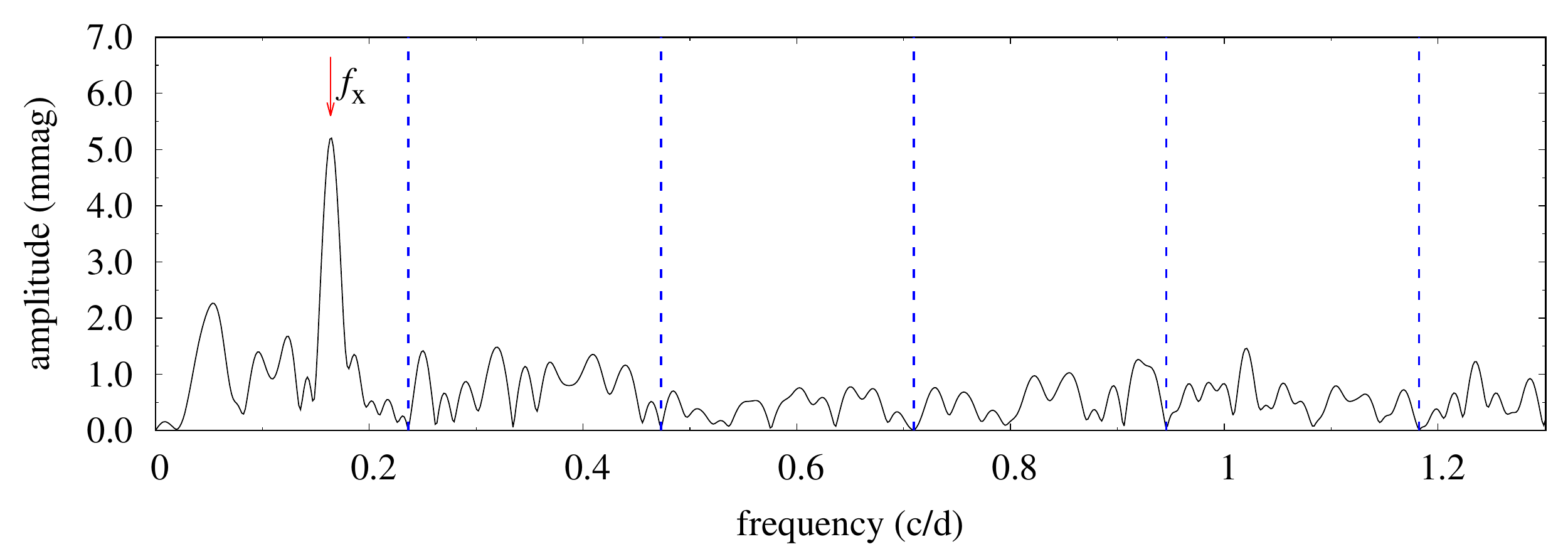}
\caption{Frequency spectrum for AH~Vel after prewhitening with the first overtone and its harmonics (dashed lines). An additional signal is marked with an arrow.}
\label{fig:ahvel}
\end{figure}

{\bf AH~Vel}. The data for AH~Vel consist of a few parts that significantly differ in quality. The mean flux level and pulsation amplitude (in flux) differ between the parts. The origin of these effects and the correct way of data processing needs to be investigated. Here we restrict the analysis to a single data chunk with the lowest dispersion of the light curve. Interestingly, after prewhitening with the first overtone and its harmonics, additional periodicity is clearly detected in the frequency spectrum (Fig.~\ref{fig:ahvel}). The period ratio, $P_{\rm 1O}/P_{\rm x}=0.693$, may indicate that additional periodicity corresponds to the radial fundamental mode. We note however, that the period ratio is a bit too low, even at such relatively long pulsation periods. Also, at such long periods, it is the fundamental mode that typically dominates the double-mode pulsation. The periodicity needs to be confirmed in other data available for AH~Vel, once these are correctly processed. 

{\bf MY~Pup}. The new data gathered during the VelPic-II campaign show complex systematics, most likely of instrumental origin, that needs to be investigated in detail and corrected. Hence, in Fig.~\ref{fig:lcO} we only show the light curve copied from our previous report (VelPic-I data).

\acknowledgements{This research is supported by the Polish National Science Centre through grants DEC-2015/17/B/ST9/03421 and 2011/01/M/ST9/05914. The operations of the Polish BRITE satellites are supported by a SPUB grant by the Polish Ministry of Science and Higher education (MNiSW). GAW is supported by an NSERC (Canada) Discovery Grant. Based on data collected by the BRITE Constellation satellite mission, designed, built, launched, operated and supported by the Austrian Research Promotion Agency (FFG), the University of Vienna, the Technical University of Graz, the Canadian Space Agency (CSA), the University of Toronto Institute for Aerospace Studies (UTIAS), the Foundation for Polish Science \& Technology (FNiTP MNiSW), and National Science Centre (NCN).}

\bibliographystyle{ptapap}
\bibliography{smolec}

\end{document}